\documentclass[conference]{IEEEtran}

\IEEEoverridecommandlockouts
\usepackage{cite}
\usepackage{amsmath,amssymb,amsfonts}
\usepackage{algorithmic}
\usepackage{graphicx}
\usepackage{parskip}
\usepackage{microtype}
\usepackage{booktabs}
\usepackage{multirow}
\usepackage{array}
\usepackage{enumitem}
\usepackage{tikz}
\usepackage{pgfplots}
\pgfplotsset{compat=1.18}
\usetikzlibrary{arrows.meta,shapes.geometric,positioning,fit,backgrounds,
                decorations.pathreplacing,calc}

\usepackage{acronym}
\usepackage{textcomp}
 \usepackage{balance}
\usepackage{wrapfig}
 \usepackage{booktabs}
\usepackage{tabularx}
\usepackage[font=footnotesize]{caption}

\usepackage{subcaption}

\usepackage{xcolor}
\usepackage{tabularray}
\usepackage{multirow}

\usepackage{eurosym}
\usepackage{tabularx}
\def\BibTeX{{\rm B\kern-.05em{\sc i\kern-.025em b}\kern-.08em
    T\kern-.1667em\lower.7ex\hbox{E}\kern-.125emX}}
\usepackage[hyphens]{url}
\usepackage[hidelinks]{hyperref}
\hypersetup{breaklinks=true}
\begin{document}
\title{Community-to-Vehicle: Integrating Electric Vehicles into Energy Communities -- A Swiss Case Study

\thanks{This research has received funding from the European Union’s Horizon Europe research and innovation programme under Grant Agreement for Project Nº 101235482 and the Swiss Secretariat for Education, Research, and Innovation (SERI) under contract Nº25.00325.}}

\author{

\IEEEauthorblockN{
Na Li\textsuperscript{1},
Dong Liu\textsuperscript{2},
Stavros Orfanoudakis\textsuperscript{2},
Özge Okur\textsuperscript{3}, N. K. Panda\textsuperscript{2,4},
Pedro P. Vergara\textsuperscript{2},
Binod Koirala\textsuperscript{1}
}

\IEEEauthorblockA{
\textsuperscript{1}Urban Energy Systems Laboratory, Empa, Swiss Federal Laboratories for Materials Science and Technology\\
Überlandstrasse 129, 8600 Dübendorf, Switzerland
}

\IEEEauthorblockA{
\textsuperscript{2}Intelligent Electrical Power Grids, Delft University of Technology, Delft, The Netherlands
}

\IEEEauthorblockA{
\textsuperscript{3}Faculty of Technology, Policy and Management, Delft University of Technology, Delft, The Netherlands
}

\IEEEauthorblockA{
\textsuperscript{4}Enexis Netbeheer B.V., 's-Hertogenbosch, The Netherlands
}
\{na.li, binod.koirala\}@empa.ch, \{D.Liu-7, S.Orfanoudakis, o.okur-1, N.K.Panda, P.P.VergaraBarrios\}@tudelft.nl}

\maketitle

\begin{abstract}
The institutional separation between local energy communities and public electric vehicle (EV) charging limits the efficient use of locally generated renewable energy. This paper introduces the concept of community-to-vehicle (C2V) as an institutional design mechanism to bridge this gap by enabling EV charging within the community boundary, where locally generated photovoltaic (PV) surplus is preferentially allocated and offered to external users at a community charging price. Building on the recently introduced local electricity community framework in Switzerland, we design scenarios that capture the transition from full separation to coordinated EV charging and evaluate their impacts on EV users and the community. The results show that C2V significantly improves local PV utilization and enhances economic performance, reducing EV charging costs relative to commercial alternatives while generating additional revenue streams for the community. These findings highlight the potential of C2V as a practical, implementable mechanism for integrating EV charging into local energy communities, providing a clear pathway for adopting coordinated community–EV interaction within existing regulatory frameworks.

\end{abstract}
\acrodef{PV}{Photovoltaic}
\acrodef{EV}{Electric Vehicle}
\acrodef{LEC}{Local Electricity Community}
\acrodef{CPO}{Charging Point Operator}
\acrodef{V2G}{Vehicle-to-Grid}
\acrodef{V2H}{Vehicle-to-Home}
\acrodef{V2B}{Vehicle-to-Building}
\acrodef{B2V2B}{Building-to-Vehicle-to-Building}
\acrodef{C2V}{Community-to-Vehicle}
\acrodef{C2V2C}{Community-to-Vehicle-to-Community}
\acrodef{DSO}{Distribution System Operator}


\section{Introduction}

\IEEEPARstart{T}{he} decarbonization of the electricity and transport sectors is generating two distinct but spatially convergent trends at the local level. On one hand, the deployment of rooftop~\ac{PV} systems, battery storage, and collective energy management schemes has accelerated the formation of local energy communities, groups of consumers and prosumers that share energy and infrastructure within a defined geographic perimeter~\cite{koirala2016energetic, li2023economic}. On the other hand, the transition to electric mobility has been accompanied by the emergence of shared~\ac{EV} services and public charging infrastructure, concentrated in residential and commercial neighborhoods where energy communities form. In Switzerland, this convergence coincides with a regulatory opening: the new~\ac{LEC} framework, effective 1~January 2026~\cite{swiss2007stromVG, swiss2008stromVV}, enables consumers and prosumers within the same distribution network area to share locally generated electricity at an agreed internal sharing price and explicitly permits third-party participants to join the community~\cite{li2026individual}.
 
Despite sharing the same physical grid area, energy communities and~\acp{EV} remain institutionally separate. Public~\acp{CPO} purchase electricity and sell to~\ac{EV} users at a price that reflects their energy procurement, grid, and margin costs. The energy community, meanwhile, manages its members under collective self-consumption agreements and exports its surplus PV generation to the grid at relatively low feed-in prices. As a result, locally generated renewable electricity and local~\ac{EV} charging demand coexist within the same distribution network area but cannot be directly coordinated. This creates a structural mismatch between physical proximity and institutional boundaries, preventing the realization of potential efficiency gains. The consequence of this separation is economically and systemically inefficient.~\ac{EV} users pay high commercial charging prices for electricity that may be locally available at lower cost, while energy communities export surplus PV generation at prices well below retail levels. The absence of a coordination mechanism, therefore, prevents both actors from capturing value that is technically and contractually achievable under existing regulatory frameworks.
 
Several works have focused on the interactions of~\acp{EV} with larger energy systems across multiple paradigms. For instance,~\ac{V2G} enables bidirectional energy services to the upstream grid but requires wholesale market participation and imposes battery degradation, which are particularly acute for small~\acp{CPO} in residential settings~\cite{tan2016integration, sagaria2025vehicle}.~\ac{V2H}~\cite{chen2020strategic} and~\ac{V2B}~\cite{he2022optimal, borge2021combined} deploy~\ac{EV} batteries behind the meter to serve a single residence or building, demonstrating strong self-consumption improvements, but are limited to a single-entity beneficiary without community scope. Other concepts, such as~\ac{B2V2B}~\cite{barone2019building, barone2020increasing, ghafoori2023electricity} and~\ac{C2V2C}~\cite{board2024community}, extend this concept by using~\acp{EV} as mobile energy carriers, physically transporting surplus electricity between buildings or communities via driving. While effective for inter-location energy balancing, these approaches require bidirectional hardware and physical~\ac{EV} transport, and they do not address the lack of an institutional mechanism to coordinate~\ac{EV} charging with LECs within an existing settlement framework. At the community level, the most recent study quantifies the techno-economic benefits of Swiss~\acp{LEC} and shows that the local sharing price is the central parameter governing benefit distribution among members~\cite{li2026individual}, a result this paper extends by examining the sharing relationship.

This paper introduces~\ac{C2V} as an institutional design framework that enables direct coordination between~\acp{LEC} and~\ac{EV} charging demand. In the proposed framework, the energy community itself acts as the charging infrastructure provider, offering~\ac{EV} users access at a community charging price and prioritizing the use of locally generated PV surplus.~\ac{EV} users, either car-sharing vehicles, office workers, or external visitors, pay well below commercial rates. The energy community earns above the feed-in tariff on surplus and recovers a small infrastructure margin on grid-sourced energy. The main contributions of this work are three-fold: (i) the institutional design of a~\ac{C2V} framework, (ii) a quantitative assessment of~\ac{LEC} and~\ac{EV} charging with institutional separation and~\ac{C2V} coordination, (iii) an explicit analysis of how community~\ac{EV} charging price influences the cost recovery for community charging infrastructure.


\section{Conceptualization and modeling framework} \label{modeling}

\subsection{From V2G to V2C: a conceptual design}
 
\begin{figure*}[!h]
    \centering
    \includegraphics[width=0.95\linewidth]{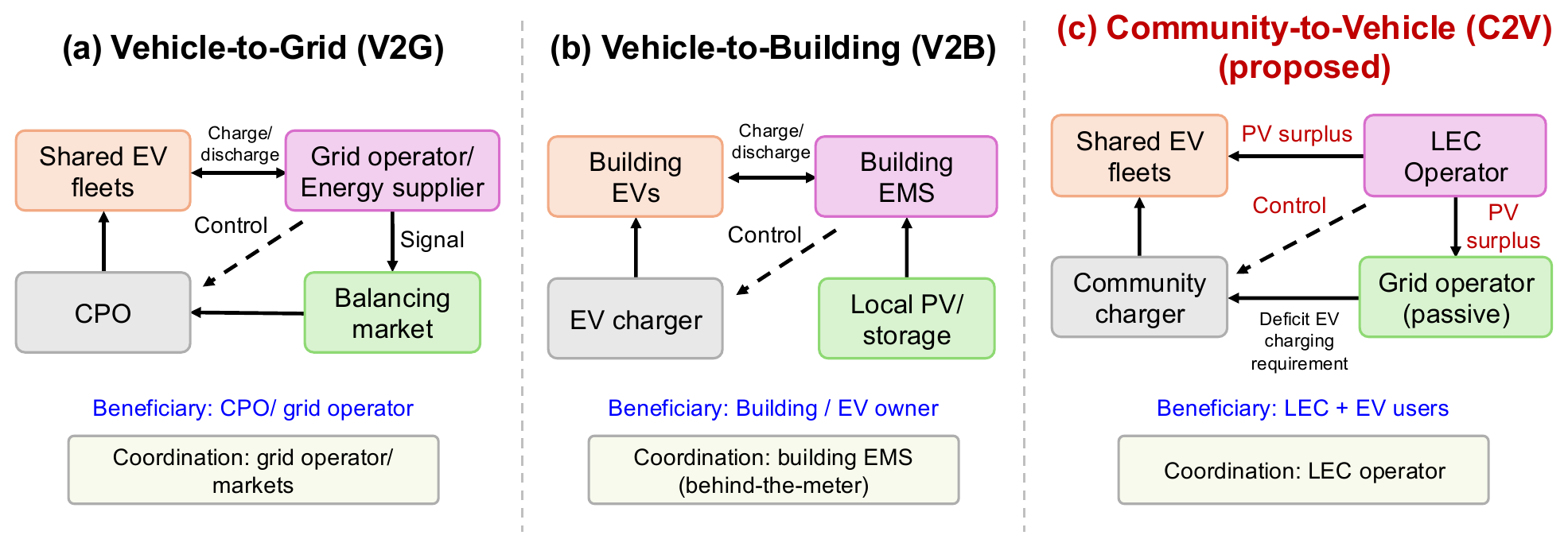}
    \caption{Paradigms of~\ac{EV} interaction with the wider electricity system. (a)~\ac{V2G}:~\ac{EV} battery serves the upstream grid; degradation compensation is required (\cite{tan2016integration}). (b)~\ac{V2B}/\ac{V2H}:~\ac{EV} serves a single building or home (\cite{he2022optimal}), with no community scope.~\ac{B2V2B} (\cite{barone2019building, barone2020increasing}) and~\ac{C2V2C} (\cite{board2024community}) use~\acp{EV} as mobile inter-location carriers, distinct from~\ac{C2V}. (c)~\ac{C2V} (proposed): the community installs charging infrastructure and directs local PV surplus to external~\ac{EV} users at the community charging price $\lambda^{\text{EV}}$. During non-surplus hours, grid energy is passed through at the same price, recovering an infrastructure margin, while~\ac{EV} users pay below commercial~\ac{CPO} rates.
}
    \label{fig:v2c_concept}
\end{figure*}

The interaction between~\acp{EV} and the wider electricity system can be understood through three distinct coordination paradigms, as illustrated in Fig.~\ref{fig:v2c_concept}. These paradigms differ primarily in their beneficiary structure and coordination layer.

In~\ac{V2G}, the~\ac{EV} is coordinated as a grid-facing asset, delivering services to upstream actors such as system operators and energy suppliers. In~\ac{V2B}/\ac{V2H}, coordination is confined to a single building or household, with benefits accruing locally but without community-level sharing. Concepts such as~\ac{B2V2B} and~\ac{C2V2C} extend the spatial scope by enabling inter-location energy transfer via mobile~\acp{EV}, but remain dependent on physical transport and specialized hardware. None of these paradigms provides a native mechanism to integrate~\ac{EV} charging into LEC settlement structures.

The~\ac{C2V} design proposed in this paper reframes the energy community itself as the charging infrastructure provider. It supplies electricity to external~\ac{EV} users, such as car-sharing vehicles, office workers, and visitors, at a community charging price $\lambda^{\text{EV}}$ that sits below the commercial~\ac{CPO} tariff $\lambda^{\text{public}}$ but above the community's retail cost of grid-sourced electricity $\lambda^{\text{buy}}$. During PV surplus hours, the community directs locally generated electricity to~\ac{EV} charging that would otherwise be exported to the grid at the feed-in tariff $\lambda^{\text{sell}}$. This allows the community to capture additional value. During non-surplus hours, grid energy is purchased at retail price and passed to external~\ac{EV} chargers at the community charging price, recovering a small infrastructure margin. This yields the price ladder:
\begin{equation}
    \lambda^{\text{sell}} <
    \lambda^{\text{loc}} <
    \lambda^{\text{buy}} <
    \lambda^{\text{EV}} <
    \lambda^{\text{public}}
    \label{eq:priceladder}
\end{equation}
where $\lambda^{\text{loc}}$ is the~\ac{LEC} internal local sharing price within the energy community.

Under institutional separation, the community exports PV surplus at low tariffs while~\ac{EV} users pay high public charging prices, leaving the intermediate value uncaptured. The proposed~\ac{C2V} framework internalizes this value within the~\ac{LEC} through two institutional elements: the community invests in charging hardware sized to the expected~\ac{EV} demand and the community's peak~\ac{PV} surplus power, and it sets a single price $\lambda^{\text{EV}}$ paid by~\ac{EV} users at all times, chosen below the commercial~\ac{CPO} rate (to attract~\ac{EV} users) and above the community's grid retail cost (to recover infrastructure and energy costs).~\ac{EV} users benefit from the price gap $(\lambda^{\text{public}} - \lambda^{\text{EV}})$ regardless of the energy source, and the community captures the remaining margin and improves the utilization of local renewable generation.

\subsection{Modeling framework}\label{sec:framework}
 
Two scenarios are compared via hourly simulation over one representative year: the institutional separation status quo (S1) and~\ac{C2V} coordination (S2).

\textit{S1 -- Institutional separation.} The~\ac{LEC} shares locally generated PV at $\lambda^{\text{loc}}$ among members~\cite{li2026individual}; surplus after local sharing is exported at $\lambda^{\text{sell}}$. External~\ac{EV} users charge independently at a commercial~\ac{CPO} at $\lambda^{\text{public}}$. The~\ac{LEC} settlement rule follows the regulatory-compliant framework formalized in our prior work~\cite{li2026individual}, and is described as follows. At each hour, each household first self-consumes its own PV generation; residual PV surplus is then shared among members with unmet demand at the internal price $\lambda^{\text{loc}}$, with network fees reduced by a factor $\gamma$ under the~\ac{LEC} regime. Unmet demand and unshared surplus exchange with the grid at $\lambda^{\text{buy}}$ and $\lambda^{\text{sell}}$, respectively.

\textit{S2 -- C2V coordination.} The~\ac{LEC} installs its own charging points, depending on the size of the surplus PV power, and applies surplus-priority dispatch: PV surplus is directed to the charging points before being exported, and residual~\ac{EV} demand is supplied through the community meter from the grid.~\ac{EV} users pay $\lambda^{\text{EV}}$ regardless of source.

In the~\ac{C2V} framework, PV surplus remaining after local sharing is directed to the community charging points before being exported. The~\ac{C2V} settlement rule is described as follows. Let $P^{\mathrm{loc}}_{t}$, $P^{\mathrm{PV2EV}}_{t}$, and $P^{\mathrm{exp}}_{t}$ denote the hourly PV flows allocated to local member sharing, PV-to-EV sharing, and grid export, respectively. The surplus-priority dispatch allocates the surplus after local sharing, $P^{\mathrm{sur}}_{t} = \sum_{i} (P^{\mathrm{gen}}_{i,t} - P^{\mathrm{self}}_{i,t}) - P^{\mathrm{loc}}_{t}$, to EV charging subject to the port-rating constraint:
\begin{equation}
P^{\mathrm{PV2EV}}_{t} =
\min\!\bigl( P^{\mathrm{sur}}_{t},\; D^{\mathrm{EV}}_{t},\;
             N_{\text{CP}} P^{\mathrm{EV,max}} \bigr),
\label{eq:share}
\end{equation}
where $D^{\mathrm{EV}}_{t}$ is the aggregate EV charging demand, $N_{\text{CP}}$ the number of charging points, and $P^{\mathrm{EV,max}}$ their rated power. Residual EV demand $P^{\mathrm{Grid2EV}}_{t} = D^{\mathrm{EV}}_{t} - P^{\mathrm{PV2EV}}_{t}$ is supplied from the grid at $\lambda^{\text{buy}}$; the unabsorbed surplus is exported, $P^{\mathrm{exp}}_{t} = P^{\mathrm{sur}}_{t} - P^{\mathrm{PV2EV}}_{t}$. In S1, $P^{\mathrm{PV2EV}}_{t} \equiv 0$.
PV-owning households receive $\lambda^{\text{loc}}$ for every unit of surplus sold locally, whether through local member sharing or through PV-to-EV sharing, allocated pro rata by their surplus contribution, so their settlement is unaffected by how the community operator allocates the surplus. 

The community charging operation is tracked in a dedicated infrastructure account. Its annual net revenue combines a surplus monetization component, the premium of  $\lambda^{\text{EV}}$ over the PV-sourced cost ($\lambda^{\text{loc}}$ paid to prosumers plus the discounted LEC network fee $(1-\gamma)\lambda^{\text{N}}$, where $\lambda^{\text{N}}$ is the volumetric network fee), and a grid pass-through margin of $\lambda^{\text{EV}}$ over $\lambda^{\text{buy}}$:
\begin{equation}
\begin{split}
R^{\mathrm{EV}} = \,&\bigl(\lambda^{\text{EV}} - \lambda^{\text{loc}} 
- (1-\gamma)\lambda^{\text{N}}\bigr) \sum_{t} P^{\mathrm{PV2EV}}_{t} 
\Delta t \\
& + \bigl(\lambda^{\text{EV}} - \lambda^{\text{buy}}\bigr) \sum_{t} 
P^{\mathrm{Grid2EV}}_{t} \Delta t,
\end{split}
\label{eq:rev}
\end{equation}
with $R^{\mathrm{EV}} = 0$ in S1. The surplus term funds the primary motivation for C2V; the pass-through term contributes to infrastructure recovery. The LEC operator uses this account for charging-point cost recovery; downstream distribution is an internal governance decision.

\subsection{Evaluation metrics}
\label{sec:metrics}


\textbf{M1 -- Local PV absorption ratio.} The share of PV generation absorbed within the community, combining household self-consumption, local member sharing, and PV-to-EV transfer:
\begin{equation}
\frac{\sum_{t} \bigl( \sum_{i} P^{\mathrm{self}}_{i,t}
                 + P^{\mathrm{loc}}_{t}
                 + P^{\mathrm{PV2EV}}_{t} \bigr)}
     {\sum_{t} \sum_{i} P^{\mathrm{gen}}_{i,t}}.
\end{equation}
It quantifies the PV surplus re-captured by~\ac{C2V} that would otherwise leave the community.

\textbf{M2 -- Grid interaction.} Community peak import $P^{\mathrm{imp}}_{\max}$ and export $P^{\mathrm{exp}}_{\max}$ capture stress at the grid connection. C2V lowers peak export by redirecting surplus to EV charging but may raise peak import for meeting residual EV charging demand.

\textbf{M3 -- Annual EV charging cost savings.} The aggregate annual savings obtained by external EV charging at the community station rather than at a commercial~\ac{CPO}:
\begin{equation}
\Delta C^{\mathrm{EV}} =
(\lambda^{\mathrm{public}} - \lambda^{\mathrm{EV}})
\sum_{t} D^{\mathrm{EV}}_{t} \Delta t.
\end{equation}
It quantifies the value transfer from commercial~\ac{CPO} revenue to external EV users under~\ac{C2V}.

\textbf{M4 -- Annual community revenue.} The annual revenue from EV charging $R^{\mathrm{EV}}$, is allocated to the community infrastructure account and decomposed into two components: (i) surplus monetization from PV-to-EV and (ii) margin from grid-supplied charging. This metric is evaluated against the charging infrastructure investment cost to assess cost recovery and payback performance.

\textbf{M5 -- PV household revenue increase.} The incremental revenue for PV-owning household $i$ arising from allocating its surplus generation to EV charging at the internal sharing price $\lambda^{\mathrm{loc}}$, rather than at the feed-in tariff: 
\begin{equation}
   \Delta R^{\mathrm{PV}}_i = (\lambda^{\mathrm{loc}} - 
\lambda^{\mathrm{sell}}) \sum_{t} P^{\mathrm{PV2EV}}_{i,t} \Delta t
\end{equation}

\section{Results and discussions} \label{Results analysis}

\subsection{Case study and input data}
\label{sec:casestudy}

The case study is based on the NEST research and innovation building at the Empa campus in D\"ubendorf, Switzerland~\cite{nest2026empa}. NEST combines residential households (renters) and office workspaces, representing a mixed-use LEC in a peri-urban Swiss context. Community demand and PV generation data are drawn from NEST metering records for 2025, comprising three PV-equipped households and four households without PV. EV charging profiles are drawn from a dataset of measured charging events at $\sim$18{,}000 public charging points across Switzerland (2022--2025)~\cite{santarromana2026swisspubliccharging}. Restricting the dataset to Zurich region in 2025 and to the 11-kW rating yields 20 representative charging profiles, each reflecting the aggregated natural mix of EV charging observed at a single public charging point.

Key parameters, including annual PV generation, annual load demand, price parameters, and EV charging information, are summarized in Table~\ref{tab:params}. To capture representative~\ac{C2V} operating conditions, we define three charging scenarios: \emph{Low}, \emph{Medium}, and \emph{High}, classified by the surplus PV-to-EV utilization ratio, corresponding to limited, moderate, and high levels of local~\ac{PV} absorption, respectively. 
\begin{table}[ht]
\caption{Input parameters}
\label{tab:params}
\centering
\footnotesize
\setlength{\tabcolsep}{2pt}
\begin{tabular}{@{}lll@{}}
\toprule
\textbf{Parameter} & \textbf{Value} & \textbf{Source} \\
\midrule
Annual community demand & 44075\,kWh & NEST\\
PV capacity & 36\,kWp & NEST \\
Annual PV generation & 22238\,kWh & NEST\\
Community charging points & 2 $\times$ 11\,kW  & Ass.\\
Grid retail tariff $\lambda^{\text{buy}}$ & 0.241\,CHF/kWh & \cite{ekz2026standardtariff}\\
Feed-in tariff $\lambda^{\text{sell}}$ & 0.06\,CHF/kWh &\cite{ekz2026feedintariff}\\
Local sharing price $\lambda^{\text{loc}}$ & 0.10\,CHF/kWh & \cite{li2026individual}\\
Community EV price $\lambda^{\text{EV}}$ & 0.30--0.55\,CHF/kWh (swept) & Ass.\\
Public charging price (AC) $\lambda^{\text{public}}$ & 0.57\,CHF/kWh &\cite{charge2026tariff}\\
Volumetric network usage fee $\lambda^{\text{N}}$ & 0.0859\,CHF/kWh &\cite{ekz2026standardtariff}\\
\bottomrule
\end{tabular}
\end{table}

\subsection{Local PV utilization and household benefits}
\label{sec:res_pv}

\begin{figure}[ht]
\centering
\includegraphics[width=\columnwidth]{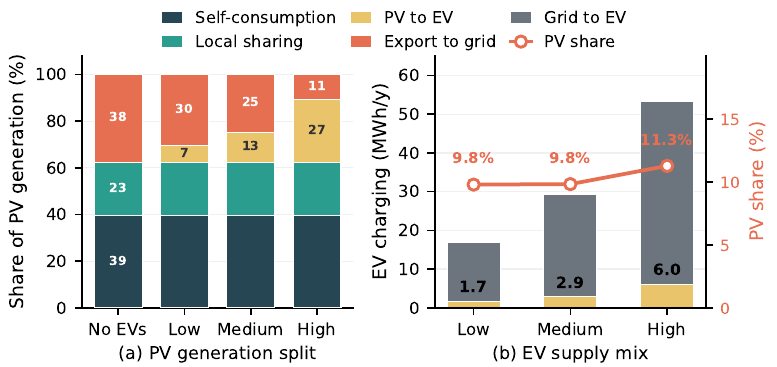}
\caption{Local PV absorption ratio decomposition: (a) PV generation split across scenarios, as percentages of annual generation. (b) EV charging supply mix: stacked bars show PV-to-EV and grid-to-EV (left axis); the line shows the share of EV charging supplied by 
community surplus PV (right axis).}
\label{fig:pv_split}
\end{figure}

Fig.~\ref{fig:pv_split}(a) decomposes the annual PV generation into self-consumption, local sharing, PV-to-EV charging, and export to the grid across the four scenarios. Self-consumption and local member sharing are determined entirely by the LEC settlement rule and remain fixed across scenarios.~\ac{C2V} operates exclusively on the residual surplus that would otherwise be exported, progressively shifting energy from export to EV charging. Accordingly, the PV local absorption rate climbs from 62\% under \emph{Low} to 89\% under \emph{High}, a 27\% gain obtained without any change to household settlement. Fig.~\ref{fig:pv_split}(b) explains the driver behind this increase: EV charging increasingly absorbs PV surplus, reaching up to 6~MWh annually under the \emph{High} scenario.

\begin{figure}[ht]
\centering
\includegraphics[width=\columnwidth]{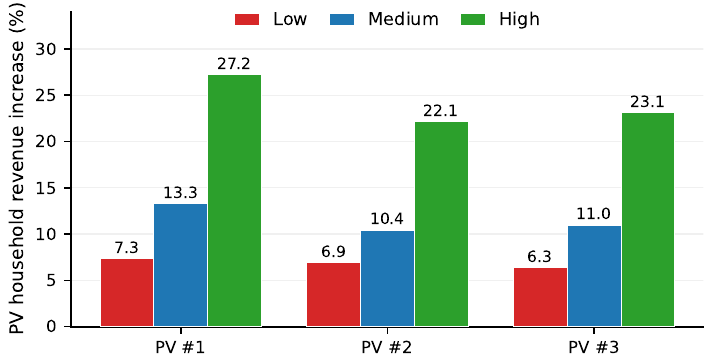}
\caption{Annual revenue increase per PV household across scenarios.}
\label{fig:pv_hh_revenue}
\end{figure}

This reallocation translates directly into PV-household revenue because redirected surplus is settled at the local sharing price rather than at the feed-in tariff. Fig.~\ref{fig:pv_hh_revenue} shows the resulting annual revenue increase for each PV household. All PV households gain in every scenario following the introduction of~\ac{C2V}. Moreover, the percentage gain is nearly identical across the three households within each scenario  due to the pro-rata settlement rule, despite differences in PV capacity.~\ac{C2V} therefore strengthens the economic case for PV households without biasing it toward large installations.

\subsection{Community and EV user benefits}
\label{sec:res_actors}

\begin{figure}[ht]
\centering
\includegraphics[width=1.05\columnwidth]{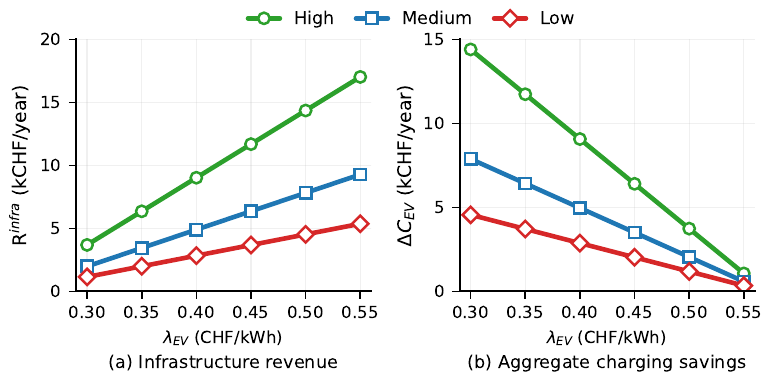}
\caption{(a) Community infrastructure revenue and (b) aggregate charging savings versus the local charging price $\lambda^{\mathrm{EV}}$ across the three scenarios.}
\label{fig:actors}
\end{figure}

Fig.~\ref{fig:actors} shows the annual community infrastructure revenue and aggregate charging savings across the full sweep of community charging prices and the three scenarios. Infrastructure revenue increases with the rise in community charging prices across the three scenarios, while charging savings move in the opposite direction. Already at a moderate $\lambda^{\mathrm{EV}}=0.40$~CHF/kWh, the annual revenue appears sufficient to recover the roughly 6~kCHF installation cost of two 11~kW AC chargers within a few years, even within a year under the \emph{High} scenario. These results expose the central design trade-off the community faces when setting the charging price: the price redistributes value only between the two parties, and the community must decide how to split it.

\subsection{Grid interaction}
\label{sec:res_grid}
A natural concern is how C2V affects the community grid connection. In our case study, peak import increases by about 20~kW (from 10.8 to 30--32~kW), matching the combined rated power of the two chargers, while peak export is unchanged because the annual surplus peak does not coincide with the selected charging profiles. This shows that C2V may or may not reduce reverse flow, but it does not worsen it. By contrast, the increase in import capacity is a direct consequence of grid-sourced charging and may require an upgrade to the grid connection before~\ac{C2V} deployment.

\subsection{Discussions}
The results show that the effectiveness of~\ac{C2V} interaction is governed by two structural elements: the LEC settlement boundary and the local charging price. By placing external EV charging within the community boundary,~\ac{C2V} reallocates residual PV surplus without interfering with existing household settlements, leading to gains in PV utilization and uniform revenue improvements across PV households.

At the community level, the local charging price does not affect physical energy flows but determines how the value created by~\ac{C2V} is distributed between PV households, EV users, and the infrastructure operator. This creates a trade-off between infrastructure cost recovery and EV user-side charging savings, which must be resolved through pricing design rather than operational changes. From a system perspective,~\ac{C2V} improves local consumption without increasing reverse power flows but shifts part of the demand to the grid through EV charging. The resulting increase in peak import is bounded by charger capacity and represents a predictable constraint rather than an operational risk.
\par

\section{Conclusions} \label{Conclusions}
This paper introduced a community-to-vehicle concept as an institutional design mechanism to integrate EV charging into local energy communities and applied it to a real-world case study under the Swiss LEC regulatory framework. C2V enables the redirection of surplus PV generation to external EV demand within the community boundary, improving local energy utilization and creating economic value without changes to household-level settlements or complex coordination. The community charging price serves as a redistribution lever between the infrastructure account and charging customers, allowing operators to balance cost recovery with user savings through pricing policy rather than operational changes. Given its low implementation barrier and system-level benefits, C2V represents a practical entry point for LEC operators to enhance local flexibility and value creation. Future work could extend the framework to include V2C discharge for peak-shaving, incorporate stochastic EV availability and PV generation, jointly optimize charging price and infrastructure sizing, and explore aggregation across multiple communities for system services and coordinated operation.

\balance

\footnotesize
\setlength{\itemsep}{0pt}
\setlength{\parskip}{0pt}
\renewcommand{\baselinestretch}{1.0}
\bibliographystyle{IEEEtran}
\bibliography{ref}

@Misc{swiss2007stromVG,
  title={Federal Law on Electricity Supply (Electricity Supply Act, StromVG)},
  organization={The Swiss Federal Council},
  year={2007},
  url={https://www.fedlex.admin.ch/eli/cc/2007/418/de}
}

@Misc{swiss2008stromVV,
  title={Electricity Supply Ordinance (StromVV)},
  organization={The Swiss Federal Council},
  year={2008},
  url={https://www.fedlex.admin.ch/eli/cc/2008/226/de}
}

@article{li2023economic,
  title={Economic analysis of energy communities: Investment options and cost allocation},
  author={Li, Na and Okur, {\"O}zge},
  journal={Applied Energy},
  volume={336},
  pages={120706},
  year={2023},
  publisher={Elsevier}
}

@misc{ekz2026standardtariff,
  title={EKZ standard tariffs for private customers 2025},
  organization={EKZ},
  year={2026},
  url={https://www.ekz.ch/dam/ekz/privatkunden/strom/tarife-und-agb/Tarifdokumente/tarife-2026/ekz-tarife-2026-privatkunden.pdf}
}

@misc{ekz2026feedintariff,
  title={EKZ feed-in tariffs for private customers 2025},
  organization={EKZ},
  year={2026},
  url={https://www.ekz.ch/dam/ekz/privatkunden/strom/tarife-und-agb/Tarifdokumente/tarife-2026/ekz-tarifsammlung-2026.pdf}
}

@misc{nest2026empa,
  title={NEST},
  organization={Empa},
  year={2026},
  url={https://www.empa.ch/web/nest/overview}
}

@misc{charge2026tariff,
  title={MOVE light charging prices},
  organization={MOVE},
  year={2026},
  url={https://www.move.ch/en/professional/subscriptions/move-light.php}
}

@article{sagaria2025vehicle,
  title={Vehicle-to-grid impact on battery degradation and estimation of {V2G} economic compensation},
  author={Sagaria, Shemin and van der Kam, Mart and Bostr{\"o}m, Tobias},
  journal={Applied Energy},
  volume={377},
  pages={124546},
  year={2025},
  publisher={Elsevier}
}

@article{borge2021combined,
  title={Combined vehicle to building ({V2B}) and vehicle to home ({V2H}) strategy to increase electric vehicle market share},
  author={Borge-Diez, David and Icaza, Daniel and A{\c{c}}{\i}kkalp, Emin and Amaris, Hortensia},
  journal={Energy},
  volume={237},
  pages={121608},
  year={2021},
  publisher={Elsevier}
}

@article{board2024community,
  title={Community-to-vehicle-to-community ({C2V2C}) for inter-community electricity delivery and sharing via electric vehicle: Performance evaluation and robustness analysis},
  author={Board, Anthony and Sun, Yongjun and Huang, Pei and Xu, Tao},
  journal={Applied Energy},
  volume={363},
  pages={123054},
  year={2024},
  publisher={Elsevier}
}

@article{ghafoori2023electricity,
  title={Electricity peak shaving for commercial buildings using machine learning and vehicle to building ({V2B}) system},
  author={Ghafoori, Mahdi and Abdallah, Moatassem and Kim, Serena},
  journal={Applied Energy},
  volume={340},
  pages={121052},
  year={2023},
  publisher={Elsevier}
}

@article{barone2020increasing,
  title={Increasing self-consumption of renewable energy through the Building to Vehicle to Building approach applied to multiple users connected in a virtual micro-grid},
  author={Barone, Giovanni and Buonomano, Annamaria and Forzano, Cesare and Giuzio, Giovanni Francesco and Palombo, Adolfo},
  journal={Renewable energy},
  volume={159},
  pages={1165--1176},
  year={2020},
  publisher={Elsevier}
}

@article{he2022optimal,
  title={Optimal integration of Vehicle to Building ({V2B}) and Building to Vehicle ({B2V}) technologies for commercial buildings},
  author={He, Zhanwei and Khazaei, Javad and Freihaut, James D},
  journal={Sustainable Energy, Grids and Networks},
  volume={32},
  pages={100921},
  year={2022},
  publisher={Elsevier}
}

@article{chen2020strategic,
  title={Strategic integration of vehicle-to-home system with home distributed photovoltaic power generation in {S}hanghai},
  author={Chen, Jianhong and Zhang, Youlang and Li, Xinzhou and Sun, Bo and Liao, Qiangqiang and Tao, Yibin and Wang, Zhiqin},
  journal={Applied Energy},
  volume={263},
  pages={114603},
  year={2020},
  publisher={Elsevier}
}

@article{barone2019building,
  title={Building to vehicle to building concept toward a novel zero energy paradigm: Modelling and case studies},
  author={Barone, Giovanni and Buonomano, Annamaria and Calise, Francesco and Forzano, Cesare and Palombo, Adolfo},
  journal={Renewable and Sustainable Energy Reviews},
  volume={101},
  pages={625--648},
  year={2019},
  publisher={Elsevier}
}

@article{koirala2016energetic,
  title={Energetic communities for community energy: A review of key issues and trends shaping integrated community energy systems},
  author={Koirala, Binod Prasad and Koliou, Elta and Friege, Jonas and Hakvoort, Rudi A and Herder, Paulien M},
  journal={Renewable and Sustainable Energy Reviews},
  volume={56},
  pages={722--744},
  year={2016},
  publisher={Elsevier}
}

@article{tan2016integration,
  title={Integration of electric vehicles in smart grid: A review on vehicle to grid technologies and optimization techniques},
  author={Tan, Kang Miao and Ramachandaramurthy, Vigna K and Yong, Jia Ying},
  journal={Renewable and Sustainable Energy Reviews},
  volume={53},
  pages={720--732},
  year={2016},
  publisher={Elsevier}
}

@dataset{santarromana2026swisspubliccharging,
  author       = {Rudolph Santarromana},
  title        = {rudolphsanta/Swiss-public-charging-dataset: Electric vehicle charging dataset on public infrastructure usage in Switzerland},
  year         = {2026},
  version      = {1.0},
  publisher    = {Zenodo},
  doi          = {10.5281/zenodo.18324722}
}

@misc{li2026individual,
      title={From Individual Consumers to Energy Communities: A Techno-economic Assessment of {Swiss} Local Electricity Communities}, 
      author={Na Li and Binod Koirala},
      year={2026},
      eprint={2604.15900},
      archivePrefix={arXiv},
      primaryClass={eess.SY},
      url={https://arxiv.org/abs/2604.15900}, 
}

\end{document}